\documentclass[sigconf, authorversion, screen]{acmart}
\AtBeginDocument{%
  }

\copyrightyear{2026}
\acmYear{2026}
\setcopyright{cc}
\setcctype{by}
\acmConference[UIST Adjunct '26]{The 39th Annual ACM Symposium on User Interface Software and Technology}{November 02--05, 2026}{Detroit, MI, USA}
\acmBooktitle{The 39th Annual ACM Symposium on User Interface Software and Technology (UIST Adjunct '26), November 02--05, 2026, Detroit, MI, USA}
\acmDOI{10.1145/3830397.3841791}
\acmISBN{979-8-4007-2855-6/2026/11}

\begin{document}

\title{Art2Song: Enhancing Visual Art Appreciation with Contextual Music Generation}

\author{Sungeun Jo}
\orcid{0009-0008-3817-2744}
\authornote{Both authors contributed equally to this research.}
\affiliation{%
  \institution{Pukyong National University}
  \city{Busan}
  \country{Republic of Korea}
}
\email{liana347@pukyong.ac.kr}

\author{Myung Jin (MJ) Kim}
\orcid{0000-0001-9970-4056}
\authornotemark[1]
\affiliation{%
  \institution{Electronics and Telecommunications Research Institute}
  \city{Daejeon}
  \country{Republic of Korea}
}
\email{mj@etri.re.kr}

\author{Chi Yoon Jeong}
\orcid{0000-0001-7089-2516}
\affiliation{%
  \institution{Electronics and Telecommunications Research Institute}
  \city{Daejeon}
  \country{Republic of Korea}
}
\email{iamready@etri.re.kr}

\renewcommand{\shortauthors}{Jo et al.}

\begin{abstract}
Art2Song is a conceptual framework that expresses artworks as sound by separating Non-Visual Context, which is difficult to perceive from the image alone, from Visual Evidence. Visual Evidence, such as objects, colors, and spatial composition, is transformed into Lyrics, while the Contextual Mood derived from the historical and art-historical context in the museum's artwork description is reflected in the background soundtrack. Rather than describing artworks textually, Art2Song aims to explore the possibility of a new mode of art appreciation in which viewers experience hidden stories and emotional context through music. As future interaction directions, we plan an Emotional Layer Blending Slider interface and a structured, traceable song-generation scenario, presenting the possibility that users can explore the relationship between Visual Evidence and Contextual Mood.
\end{abstract}



\begin{teaserfigure}
 \includegraphics[width=\textwidth]{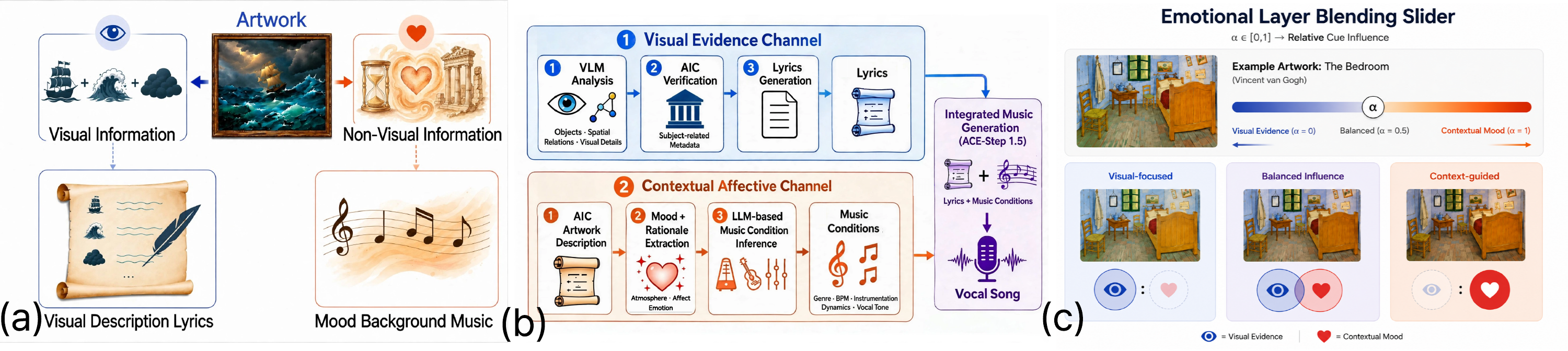}
  \caption{Art2Song Overview. (a) Core concept: separating a painting's Visual Evidence and Contextual Mood into distinct song components (Lyrics and background soundtrack). (b) Prototype pipeline structure for artwork-to-song generation. (c) Proposed concept of modulating the influence of Visual Evidence and Contextual Mood on song generation (the Emotional Layer Blending Slider).}
  \Description{A three-panel overview diagram. Panel (a) shows a painting split into two streams: visual information feeding the Lyrics and contextual information feeding the background soundtrack. Panel (b) shows the prototype pipeline from an artwork image and Art Institute of Chicago metadata through vision-language model and language model stages to ACE-Step music generation, and panel (c) shows a slider ranging between Visual Evidence and Contextual Mood endpoints.}
  \label{fig:teaser}
\end{teaserfigure}


\maketitle

\section{Introduction}
Although art appreciation is primarily visual, historical background and art-historical interpretation often deepen it. This work explores whether such visually hidden contexts can be conveyed through sound.
Museum and HCI research treats musical augmentation as interpretation, not neutral translation: viewers connect paintings and music through emotion, movement, narrative, and theme; curated pairings can feel more meaningful than automatic ones~\cite{soundofpaintings2023,warren2025crossmodal}. Multisensory experiences can also influence or shape how artworks are interpreted~\cite{vi2017tateSensorium}; systems should make their interpretive stance visible and support alternative readings.
Image-to-Music systems translate visual features into music~\cite{Wang2025VisionToMusic, chowdhury2023melfusion}, and recent approaches jointly condition generation on visual and contextual information~\cite{art2music2025, art2mus2026, xin2025semantic}. While effective, such designs do not necessarily make explicit how each source shapes the resulting interpretation. Van Gogh’s \textit{The Bedroom} appears calm, yet the museum’s description interprets it as expressing longing for rest and inner tension.
Art2Song attends to this gap, assigning visible and non-visible information distinct auditory roles reflecting expressive affordance: language suits explicit description, so directly observable, verbalizable objects, colors, and spatial relations become Lyrics; music can convey affect nonverbally, so the Contextual Mood from the work’s historical and interpretive context becomes a background soundtrack shaping the song’s overall atmosphere. Art2Song proposes appreciating the scene before the viewer and the context behind it together within a single song.

\section{Proposed Concept Prototype Pipeline}
Our initial prototype comprises a Visual Evidence Channel conveying observable visual information through Lyrics, and a Contextual Affective Channel conveying the Contextual Mood derived from the artwork's non-visual context through the background soundtrack. The Art Institute of Chicago (AIC) API~\cite{aicapi} was selected as the initial data source because it provides subject-related metadata and curated artwork descriptions through a single API. Title, accession number, artist, and date identify the artwork; subject metadata verifies extracted objects (Visual Evidence Channel); the description's historical and art-historical context grounds Contextual Mood extraction (Contextual Affective Channel).

In the Visual Evidence Channel, a vision-language model (VLM; Qwen3.5~\cite{qwen3.5}) extracts objects, positions, spatial relations, colors, and scene composition from the artwork image, excluding symbolic and historical interpretation. Since AIC subject metadata lists only main subjects and objects, the VLM adds the spatial detail Lyrics require (e.g., in \textit{The Bedroom}, the bed on the right, table on the left, window above the table). The metadata also cross-checks the VLM output, marking provenance: matched objects are cross-confirmed; unmatched objects are retained but marked as image-only evidence. The organized information becomes general-to-specific Lyrics.

In the Contextual Affective Channel, Qwen3.5 extracts the Contextual Mood and a recorded rationale from the artwork description, grounded in curatorial and art-historical interpretation and the artist's statements where provided. Both are fed back to Qwen3.5 to produce structured music-generation conditions: genre, BPM, instrumentation, dynamics, and vocal tone. Emotional cues are not mapped one-to-one to musical parameters; the model selects them within a fixed output format and value ranges, jointly considering multiple contextual cues. For \textit{The Bedroom}, the extracted Contextual Mood included repose, nervous energy, instability, turmoil, and complexity, leading to an ambient, minimalist, atmospheric genre at 72 BPM with piano, strings, and soft pads, quiet-to-swelling tense dynamics, and a breathy, calm, slightly anxious vocal tone.

The two channels' Lyrics and music-generation conditions are fed together into ACE-Step 1.5~\cite{acestep1.5-2026} to generate an approximately 60-second vocal song. The generated examples showed that Lyrics carrying visual content and a background soundtrack reflecting the Contextual Mood can be in harmony or in contrast. This suggests a design possibility of examining visual impression and context-derived emotion separately. Sample audio files and metadata are provided as supplementary material.

Several failure types were observed: the VLM sometimes did not fully follow extraction constraints, omitting requested visual elements or producing abstract expressions instead of observable descriptions, and parts of the Lyrics were dropped or rendered unclearly during music generation. Future work will systematically evaluate their frequency and improve the model (e.g., fine-tuning) and generation conditions so intended visual information and Lyrics are reflected consistently.

\section{Expected Experience and Future Interaction}
\subsection{New Way of Appreciating Art}
Art2Song explores the possibility of adding a new experiential layer to art appreciation rather than replacing text-based artwork descriptions. Users listen to vocal music: the general-to-specific Lyrics direct attention to visible details (objects, positions, spatial relationships), while the background soundtrack conveys the Contextual Mood derived from non-visual contextual information. The structure is expected to guide users to revisit visual details they may have initially overlooked. In \textit{The Bedroom}, the Lyrics trace the room's composition while the soundtrack creates a melancholic, unstable atmosphere reflecting longing for rest and inner tension. Through this contrast, the design is intended to support viewers in considering the artwork not simply as a peaceful bedroom scene, but as an emotionally complex work shaped by its historical context. Art2Song proposes that what is visible may not be all there is to a work, exploring whether this contrast can spark curiosity about its context and further exploration.

\subsection{Emotional Layer Blending Slider}
As a future design direction, we propose a conceptual, not-yet-implemented slider for exploring the relative influence of Visual Evidence and Contextual Mood; its value $\alpha \in [0,1]$ is specified to Qwen3.5 as a contextual-influence weight on visual-cue selection (Lyrics) and contextual conditioning (music). Visual-focused ($\alpha \approx 0$) prioritizes visual relevance; Balanced Influence ($\alpha \approx 0.5$) considers both jointly; Context-guided ($\alpha \approx 1$) weighs the AIC description's Contextual Mood and rationale more heavily. Every position limits the Lyrics to directly observable Visual Evidence; the Contextual Mood re-prioritizes emphasis without adding non-visual content---e.g., at Context-guided, \textit{The Bedroom}'s bed (restfulness) and tilted composition (instability) gain prominence. Rather than crossfading audio, weighted conditions regenerate the song with ACE-Step~1.5; continuous weighting and parameter mapping remain future work.

\subsection{Structured and Traceable Song Generation}
Art2Song structures and records the evidence and intermediate representations behind the Lyrics and background soundtrack. For the Lyrics: the image-extracted objects, positions, and spatial relations cross-checked against AIC information, so their reflection can be confirmed. For the soundtrack: the AIC description with the extracted Contextual Mood and rationale, so both the inferred emotion and the music-generation conditions it led to can be traced. These records could extend into a process-visualization interface tracing artwork information into Lyrics and music, or the music back to the context that grounded it.

\section{Conclusion}
This study proposed Art2Song, a framework that conveys the visual elements and hidden contexts of artworks through Lyrics and a background soundtrack, respectively. The initial prototype suggested the potential of this visual–contextual separation while revealing challenges in visual information extraction, Lyrics preservation, and generation control.

Future work will improve pipeline reliability, explore alternative auditory formats (vocal songs versus narration-based soundtracks), and extend Art2Song through additional museum APIs and public metadata sources. User studies will compare Visual Evidence-focused, Contextual Mood-focused, and Combined settings on artwork understanding, engagement, and interpretive exploration. Art2Song aims to support interactive art appreciation in which viewers explore how visible scenes and hidden contexts relate through music.

 \begin{acks}
This work was supported by the Electronics and Telecommunications Research Institute (ETRI) grant funded by the Korean government (26ZR1200, Research on Autonomous Vision Augmentation and Extension Technologies; 25YR1900, Neuroplasticity-Based Adaptive Control Interfaces for the Embodiment of Supernumerary Artificial Limbs).
\end{acks}

\bibliographystyle{ACM-Reference-Format}
\bibliography{references_list}










\end{document}